\documentclass[epj,final]{svjour_1}
\usepackage{graphicx}
\usepackage{subfig}
\usepackage{epstopdf}
\usepackage{floatrow}
\usepackage{cite}

\begin{document}

\title{Selective excitations of a Kerr-nonlinear resonator: exactly solvable approach}
\author{
A R Shahinyan \inst{1} \and A R Tamazyan \inst{1,2} \and G Yu Kryuchkyan \inst{1,3}
}
\institute{ 
Institute for Physical Researches,National Academy of Sciences, Ashtarak-2, 0203, Ashtarak, Armenia 
\and Physics Department, Yerevan State University, Alex Manoogian 1, 0025, Yerevan, Armenia 
\and Centre of Quantum Technologies and New Materials, Yerevan State University, Alex Manoogian 1, 0025, Yerevan, Armenia
}
\date{}
%
\abstract{
We study Kerr nonlinear resonators (KNR) driven by a continuous wave field in quantum  regimes where strong Kerr interactions give rise to selective resonant excitations of oscillatory modes. We use an exact quantum theory of KNR in the framework of   the Fokker-Planck equation without any quantum state truncation or perturbation procedure.  This approach allows non-perturbative consideration of KNR for various quantum operational regimes  including   cascaded processes between oscillatory states. We focus on understanding of multi-photon non-resonant and selective resonant excitations of introcavity mode depending on the detuning, the amplitude of the driving field and the strength of  nonlinearity. The analysis is provided  on the base of photon number distributions,  the photon-number correlation function and the Wigner function.
}
\maketitle
\section{Introduction}\label{intro}

It is known that a cavity  mode strongly coupled to a nonlinear element  leading to direct photon-photon interactions behaves like a one-mode oscillator with an anharmonic energy ladder.  Large nonlinearities make the photon energy spectrum non-equidistant and  well resolved, therefore spectroscopic selective excitations of transitions between Fock states become possible. Currently existing and recently proposed oscillatory systems  realizing photon-photon interactions contain a large number of various nonlinearities including Kerr-nonlinearity. A Kerr nonlinear resonator (KNR) can serve as an example. Selective excitations of KNR can be realized by tuning the frequency of the driving field to the sum of the resonator frequency and the Kerr nonlinear shift of the energy levels. The anharmonic, Jaynes-Cummings ladder has been recently probed in atomic  \cite{kub} and superconducting  \cite{lev} cavity quantum electrodynamics (cQED) systems. It has been observed that non-classical correlations between photons being transmitted through the cavity can result from such anharmonicities,  leading to the fundamental phenomena of photon blockade (PB) and photon-induced tunneling.  In PB, the capture of a single photon by the system affects the probability of a second photon being admitted. PB has been proposed for many applications, including the generation of single photons, quantum communications, where strong antibunching of photons and sub-Poissonian statistics is required, high precision sensing and quantum metrology.

 Originally, PB was proposed in a Kerr-type nonlinear cavity via giant photon-photon interactions \cite{Imam} and was first observed in an optical cavity coupled to a single trapped atom \cite{Birn}. PB has also been predicted in cQED \cite{Tia}, and recently in circuit QED with a single superconducting artificial atom coupled to a microwave transmission line resonator \cite{Hof, Lan}. Going further, PB was experimentally demonstrated for a photonic crystal cavity with a strongly coupled quantum dot \cite{Far2}, and was  predicted in quantum optomechanical systems \cite{Rab, Nun}. An analogous phenomenon of phonon blockade was predicted for an artificial superconducting atom coupled to a nanomechanical resonator \cite{Liu}, and the polariton blockade effect due to polariton-polariton interactions has also been considered \cite{Ver}. Recently, one-photon PB was considered for dispersive qubit-field interactions in a superconductive coplanar waveguide cavity \cite{hof2}, with time-modulated input \cite{A}, and for the train of pulses \cite{hov1,hov2}. Two-photon blockades as well have been  investigated \cite{hov1,mir}.  It is commonly believed that photon blockade necessarily requires a strong  Kerr nonlinearity  for a photonic mode, magnitude of which should well exceed the mode decay rate. It has been also shown that photon blockade could be achieved in nanostructured cavities  with third-order nonlinear susceptibility  \cite{fer}. An alternative unconventional photon-blockade effect was recently proposed with a weaker Kerr nonlinearity than its conventional counterpart in a system consisting of two coupled photonic cavities or from suitably engineered coupled modes in one cavity \cite{lie}. The strong photon-photon correlation in this scheme is attributed by  the destructive quantum  interference between distinct driven-dissipative pathways \cite{car1, bam}. Based on this mechanism, many different systems are proposed to realize the  unconventional photon-blockade \cite{xu,maj,zhang,xu2,shel,fer2,shen,fay}, moreover three oscillatory mode systems have also been considered \cite{ger}.

In all cited papers, the regimes of PB  in cavity and determination of optimal values of the  detuning and the nonlinear coupling constants are analyzed in  a simplified approach based on the wave function study. Here, mode losses are treated at the non-hermitian hamiltonian level and  the wave function expands on a Fock-state basis. Assuming weak optical pumping  the  dynamics of the lowest Fock state is only analyzed in a steady state. In this way,  the master-equation solutions are obtained through  matrix elements of Fock states analytically as well as numerically for the lowest states in steady state regimes. This approximation  of Fock state truncation in a steady-state allows to consider a Kerr-type nonlinear resonator or a system of coupled resonators for the case of sufficiently weak driving external fields. In this regime the photon-number distributions are also discussed \cite{leon, kry}. Besides, the processes of  re-excitation and cascaded decay of oscillatory states are neglected in this approximation. This approach as well as the nonlinear quantum scissor effects were discussed earlier \cite{leon1}(see, also the review paper \cite{leon2}. Photon-number effects, multi- photon blockade and generation of Fock states in KNR have been studied numerically in the full time-evolution of the master equation on the base of the quantum state diffusion method \cite{gev1, gev2, gev3, gev4}.

In this paper, we analyze selective excitations of photon-number states in KNR driven by a continuous wave field in a steady state regime for strong Kerr nonlinearities. The novelty here is that we analyze selective excitations in KNR using exact quantum treatment of driven dissipative KNR in the framework of the Fokker-Planck equation  in complex P-representation \cite{drum, a33, a34, kh, crit}.  This approach gives a full quantum description of the Kerr dissipative resonator for quantum noises of arbitrary strength without any state truncation procedure  and hence allows us to find fundamental limits of PB due to dissipative and decoherence effects. Moreover, the approach allows non-perturbative consideration of KNR for various quantum operational ranges  including cascaded processes between oscillatory states and multi-photon selective resonant excitations.  The last case is realized  by tuning the frequency of the driving field for wide ranges of the parameters. We clarify the selective excitations of the oscillatory mode  leading to multi-photon blockades considering photon-number effects on the base of the mean photon number, the probability distributions of photon numbers, the second-order correlation function of photons and the Wigner function of the cavity mode.

The important parameter responsible for  KNR is the ratio $\chi/\gamma$ between the parameter of the Kerr-type nonlinearity and the damping of the oscillatory mode (the photon decay rate).  As a rule, the efficiency of quantum nonlinear effects requires a high nonlinearity with respect to the dissipation, posing a severe technological challenge. Nevertheless, resonators with Josephson junctions can  be used to reach strong nonlinear regimes. Particularly, the regimes of  weak $\chi/\gamma<<1$, strong $\chi/\gamma>1$ and very strong $\chi/\gamma>>1$ nonlinearities have been demonstrated for a Josephson junction embedded in a transmission-line resonator by adjusting the parameters of the circuit \cite{blais, blais0, blais1}. 

The paper is arranged as follows. In Sect. \ref{model} we present the quantities of interest for  KNR  by using the potential solution of the Fokker-Planck equation in complex P-representation.  In Sect. \ref{SecPD} we analyze multi-photon excitations in KNR on the base of the mean photon number,  the photon-number state populations, the second-order photon correlation function and the Wigner functions.  This analysis includes a study of  non-resonant cascaded excitations  as well as selective resonant excitations of KNR depending on the detuning, the amplitude   of driving field and the strength of the Kerr nonlinearity.  We summarize our results in Sect. \ref{Conclusion}.

\section{Exact quantum treatment of KNR:  short description }\label{model}

The Hamiltonian of Kerr-nonlinear resonator driven by a monochromatic field in the rotating-wave approximation reads as:

\begin{equation}
H_{0}=\hbar \Delta a^{\dagger}a + \hbar \chi a^{\dagger2}a^{2} +
\hbar (\Omega a^{\dagger} + \Omega^{*}a) ,\label{hamiltonian}
\end{equation}

where $a^{\dagger}$, $a$ are the oscillatory creation and annihilation operators,
$\chi$ is the nonlinearity strength and $\Delta=\omega_{0} -\omega$ is the
detuning between  the oscillatory frequency  and the frequency of the driving field.
The  Hamiltonian (\ref{hamiltonian}) describing  an anharmonic oscillator has been proposed for   wide range of physical systems, including  optical fibers,  nano-mechanical oscillators, various  Josephson junction based devices, quantum dots,  quantum scissors, etc.

 The full Hamiltonian of the system reads as  $H=H_{0}+H_{\textrm{loss}}$, where the term $H_{\textrm{loss}}=a \Gamma^{+} + a^{+}\Gamma $ is responsible for linear losses of the oscillatory mode, due to couplings with the heat reservoir operators giving rise to the damping rate  $\gamma$. We trace out the reservoir degrees of freedom in the Born-Markov limit assuming that the system and the environment are uncorrelated at initial time t = 0. This procedure leads to a master equation for the reduced density matrix of the oscillatory mode  in the following form:

\begin{equation}
\frac{\textrm{d} \rho}{\textrm{d} t} =-\frac{\textrm{i}}{\hbar}[H_{0}, \rho] +
\sum_{i=1,2} L_{i}\rho
L_{i}^{+}-\frac{1}{2}L_{i}^{+}L_{i}\rho-\frac{1}{2}\rho L_{i}^{+}
L_{i}.\label{master}
\end{equation}

Here $L_{1}=\sqrt{(N+1)\gamma}a$ and $L_{2}=\sqrt{N\gamma}a^+$ are the Lindblad operators, $\gamma$ is the dissipation rate and $N$ denotes the mean number of quanta of the heath bath.To study the pure quantum effects we focus on the cases of very low reservoir temperatures with $N$  approximately equal to zero. The master equation is then transformed into a Fokker-Planck equation. In this simplest case analytical results for the Kerr-nonlinear dissipative resonator in a steady state have been obtained in terms of the solution of the Fokker-Planck equation for the quasi-probability distribution function $P(\alpha, \alpha^{*})$ in the complex P-representation \cite{drum}. This approach  based on the method of potential equations leads to the analytic solution for the quasi-probability distribution function $P(\alpha, \alpha^{*})$ within the framework of the exact nonlinear treatment of quantum fluctuations. On the whole, the various moments of mode operators as well as the photon distribution function were obtained \cite{a33, a34, kh}. For instance, the photon number probability distribution function $P_{n} = \langle n |\rho| n \rangle$ can be expressed in terms of the complex P-representation as follows

\begin{equation}
\label{pnPc}
P_{n}=\frac{1}{n}\int \int_C \textrm {d}\alpha \textrm{d}\alpha^{*} \textrm{e}^{-\alpha \alpha^{*}}\alpha^n {\alpha^{*}}^nP(\alpha, \alpha^{*}),
\end{equation}

where C is an appropriate integration contour for each of the variables $\alpha$ and $\alpha{*}$, in individual complex planes. The details of calculations involving contour integrations have been presented in \cite{a33, a34, kh}. The results for the mean photon number and the photon number probability distribution function read as

\begin{equation}
P_{n} =\frac{|\varepsilon|^{2n}\Gamma(\lambda)\Gamma(\lambda^{*})}{n!_{0}\textrm{F}_2(\lambda; \lambda^{*}; 2 |\varepsilon|^{2})}
\sum_{k = 0}^{\infty}\frac{|\varepsilon|^{2k}}{k!\Gamma(k+n+\lambda)\Gamma(k+n+\lambda^{*})},
\label{phdis}\\
\end{equation}
\begin{equation}
\langle n\rangle=\langle a^{\dagger}a\rangle = \frac{\Omega^2}{(\Delta + \chi)^2 + (\gamma/2)^2}
\frac{{}_0\textrm{F}_{2}(\lambda + 1; \lambda^{*} + 1; z)}{{}_0\textrm{F}_{2}(\lambda;\lambda^{*}; 2 |\varepsilon|^{2}|)},
\label{quant}
\end{equation}

where $\varepsilon = \Omega/\chi$, $\lambda = (\gamma + \textrm{i}\Delta)/\textrm{i}\chi$, and ${}_{0}\textrm{F}_{2}$ is the hypergeometric function

\begin{equation}
_{0}\textrm{F}_{2}(a; b; z)=\sum_{k = 0}^{\infty}\frac{z^{k}\Gamma(a)\Gamma(b)}{k!\Gamma(k+a)\Gamma(k+b)}.
\end{equation}

In this approach, the second-order correlation function of photon-numbers for zero-delay time

\begin{equation}
g^{(2)}(0) = \frac{<a^{\dagger} a^{\dagger} a a> }{<a^{\dagger} a>^2}
\label{g_init}
\end{equation}

 is calculated in the following form

\begin{equation}
g^{(2)}(0) = \frac{|\varepsilon|^{4}\Gamma(\lambda)\Gamma({\lambda^{*}}){}_0\textrm{F}_2(2+\lambda; 2+\lambda^{*}; 2|\varepsilon|^2)}{\Gamma(2+\lambda)\Gamma(2+\lambda^{*}){}_0\textrm{F}_2(\lambda; \lambda^{*}; 2|\varepsilon|^2)}\frac{1}{\langle n\rangle^2}.
\label{g_2}
\end{equation}

Of particular interest is the analysis of phase-space properties of cavity mode in the framework of the Wigner function  derived from  the Fokker-Planck equation of KNR in a steady-state regime. The Wigner function involving quantum noises with terms of all orders of the perturbation theory reads as

\begin{equation}
W(\alpha)=N\exp^{-2|\alpha|^{2}}\left|\frac{J_{\lambda-1}(\sqrt{-8\alpha\varepsilon})}{(\alpha^{*})^{(\lambda-1)/2}}\right|^{2}.
\end{equation}

Here $J_{\lambda}$ is the Bessel function, $\lambda=(\gamma+i\Delta)/\chi$, $\varepsilon=\Omega/\chi$, $N$ is the normalized constant and the amplitude $\alpha=x+iy$ is a complex c-number variable corresponding to the operator $a$. For further calculations we use the representation of Bessel function in terms of Gamma functions and the hypergeometric function

\begin{equation}
J_{\nu}(z)=\frac{(\frac{1}{2}z)^\nu}{\Gamma(\nu+1)}{}_0\textrm{F}_1(;\nu+1; - \frac{1}{4}z^2).
\end{equation}

where,

\begin{equation}
{}_0\textrm{F}_1(;b; z)=\sum_{k = 0}^{\infty}\frac{z^{k}\Gamma(b)}{k!\Gamma(k+b)}.
\end{equation}

We then calculate  the Wigner function in the following form

\begin{equation}
W(\alpha)=N\exp^{-2\alpha^2}\Bigg|(-4\varepsilon)^\frac{\lambda-1}{2}\sum\frac{(2\alpha\varepsilon)^k}{\Gamma(\lambda+k)k!}\Bigg|^2.
\end{equation}

This approach is applied for quantum  operational regimes of KNR consisting of non-resonant and resonant excitations of the cavity mode by an external coherent field below. 

\begin{figure}%
\subfloat[][\label{pn1a}]{\includegraphics[scale=0.6]{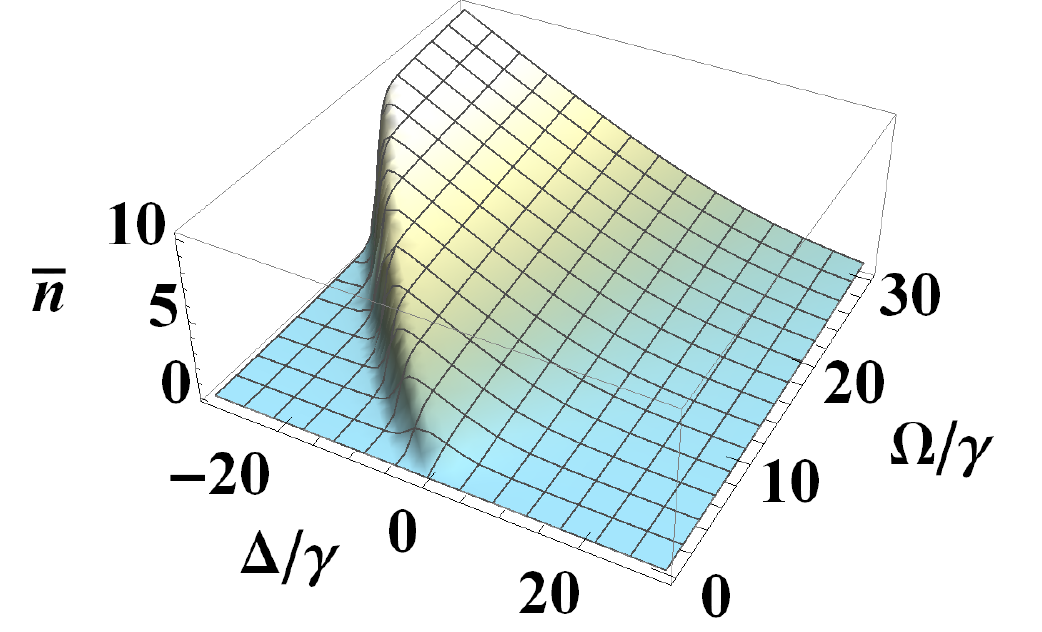}}%
\qquad
\subfloat[][\label{pn1b}]{\includegraphics[scale=0.6]{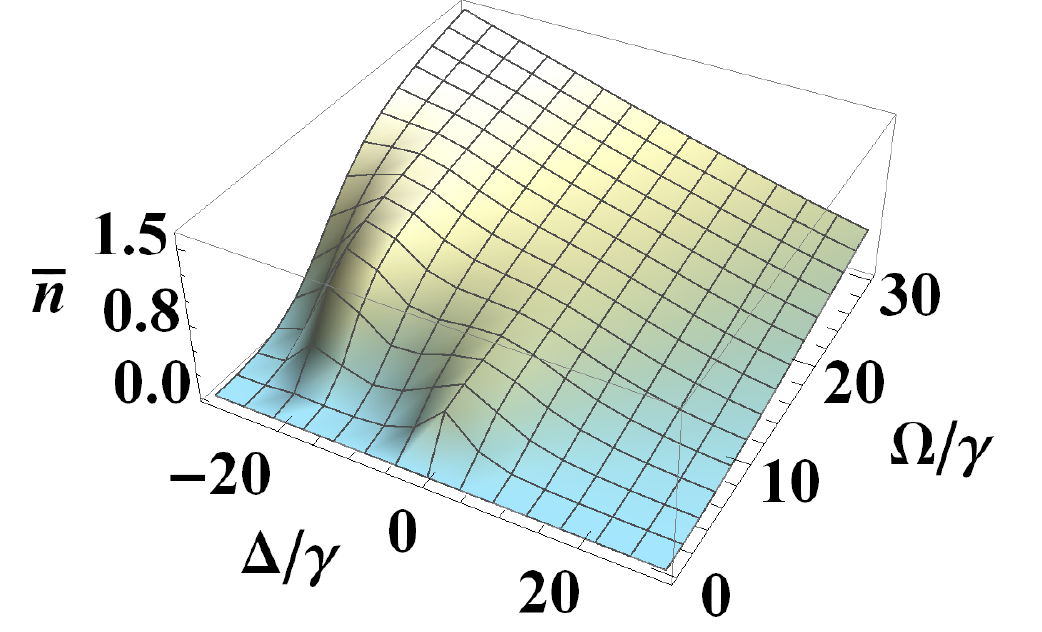}}
\caption{Photon numbers in dependence on the  detuning and the driving field
amplitude. The parameters are as follow: $\chi/\gamma=2$ (a) and $\chi/\gamma=20$ (b).}
\label{pn1}
\end{figure}

\section{Multi-photon excitations in a Kerr dissipative  resonator }\label{SecPD}

\begin{figure}%
\subfloat[][\label{pn3d2a}]
{\includegraphics[scale=0.6]{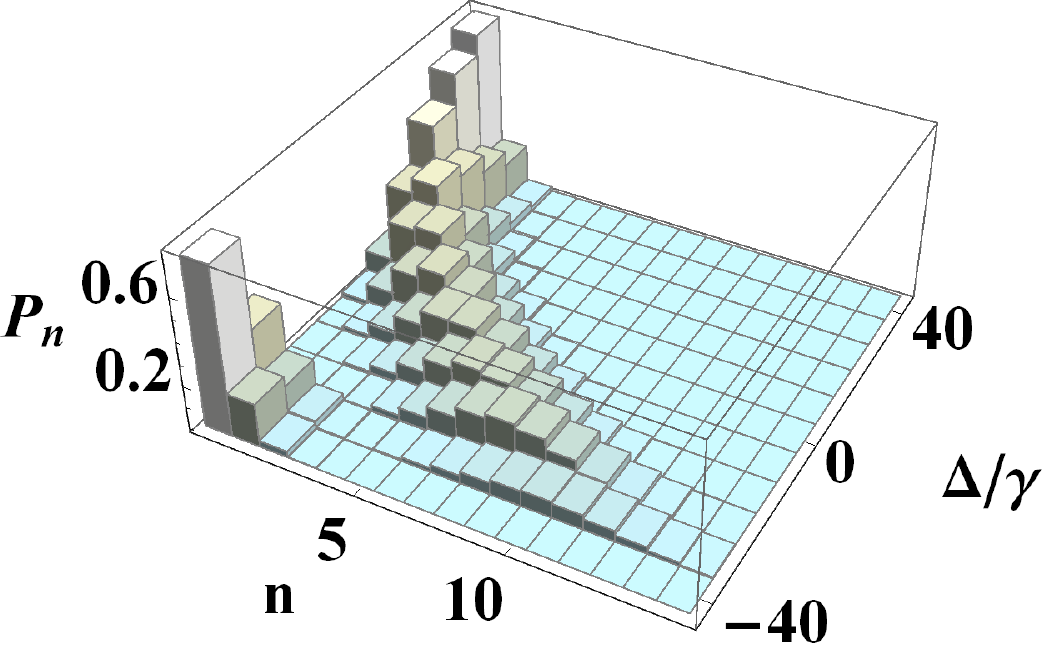}}%
\qquad
\subfloat[][\label{pn3d2b}]
{\includegraphics[scale=0.6]{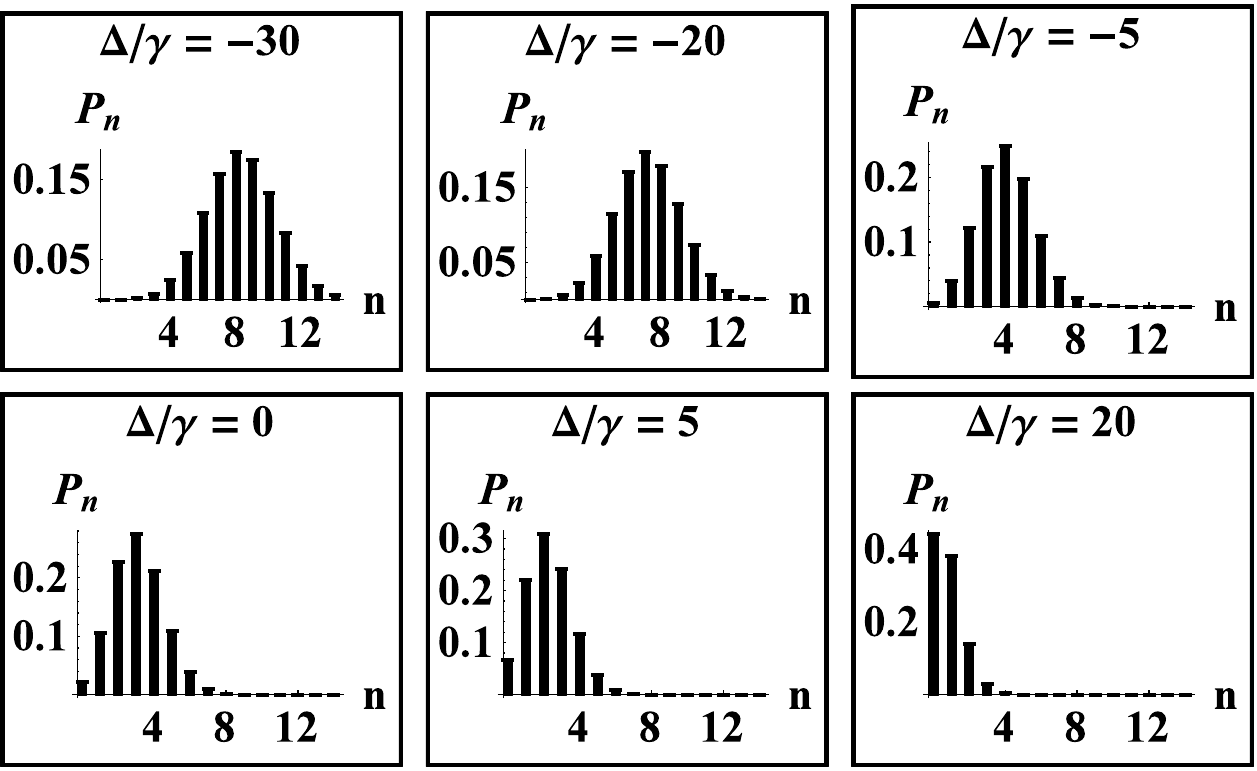}}
\caption{(Color online) Populations of photon numbers depending on the detuning (a); cross-sections of the plot for particular detuning values (b).}
\label{pn3d2}
\end{figure}

In the absence of any driving,  states of the nonlinear oscillator are photon-number states $|n\rangle$ which are spaced in energy $E_{n} = E_{0} +\hbar\omega_{0} n + \hbar\chi n(n-1)$ with $n = 0, 1, ...$. The levels form an anharmonic ladder with a nonlinear parameter that is given by $E_{21}-E_{10}=2\hbar\chi$. The energy spectrum can be probed with response of the system to the driving field when the driving frequency is tuned.  In this case, spectral lines of the system are resolved if the nonlinear shifts of oscillatory energy levels are larger than the state line-widths, i.e. $\chi/\gamma >1$. According to the formula of the energy spectrum resonant frequencies $ n\hbar \omega_n=E_{n0}$ of the multiphoton transitions  $|0\rangle\rightarrow|n\rangle$ are derived in the following form $\omega_n=\omega_0+\chi (n-1)$. Thus, in order to excite a state with one photon ($E_{10}=\hbar \omega_0$) the driving frequency is fixed to the resonant frequency $\omega_1=\omega_0$. For the two-photon transition $|0\rangle\rightarrow|2\rangle$, $E_{20}=2 \hbar \omega_0 + 2 \chi $, and  the resonant frequency is $\omega_2=\omega_0+ \chi $, while for the three-photon transition $|0\rangle\rightarrow|3\rangle$,  $E_{30}=3 \hbar \omega_0 + 6 \chi$, the resonant frequency is $\omega_3=\omega_0+ 2\chi $. Hence, the detuning values at the resonant frequencies can be written as $\Delta_{n}=\omega_{0}-\omega_{n}=-\chi (n-1)$. This analysis is correct if the driving field amplitude is weak. Otherwise  the spectrum of mode excitations  will contains  frequencies corresponding to oscillatory Raman processes under the external drive, moreover, the resonant frequencies will be  shifted  due to the Stark effects as well.  Indeed, in the second-order approximation of the perturbation theory for the interaction of an anharmonic oscillator with a monochromatic field, the shift of oscillatory energy $E_{n}$ in transitions through the states $|n-1\rangle$ and $|n+1\rangle$  is obtained as $\Delta E_{n} = \hbar\Omega^{2} \left(\frac{n}{\omega + \chi(2n-1)} - \frac{(n + 1)}{\omega + \chi(2n+1)}\right)$.

 As the calculations based on Eqs. (\ref{phdis}), (\ref{quant}) and (\ref{g_2}) show, the  results strongly depend on all the parameters: the nonlinearity, the detuning and the amplitude of the driving field. Below, we focus on comparative analysis of the regimes of strong and very strong nonlinearities  in order to demonstrate selective excitations in KNR due to formation of  well-resolved multi-photon resonance spectral lines. In this way, we concentrate on the particular cases with  the values of nonlinearity: $\chi/\gamma=2$ and  $\chi/\gamma=20$, considering the interplay of the ratio $\chi/\gamma$,  the detuning $\Delta/\gamma$ and the amplitude of driving field  $\Omega/\gamma$.

\subsection{Non-selective excitations of oscillatory states}\label{SecPD1}

\begin{figure}%
\subfloat[][\label{pdelta3a}]{\includegraphics[scale=0.35]{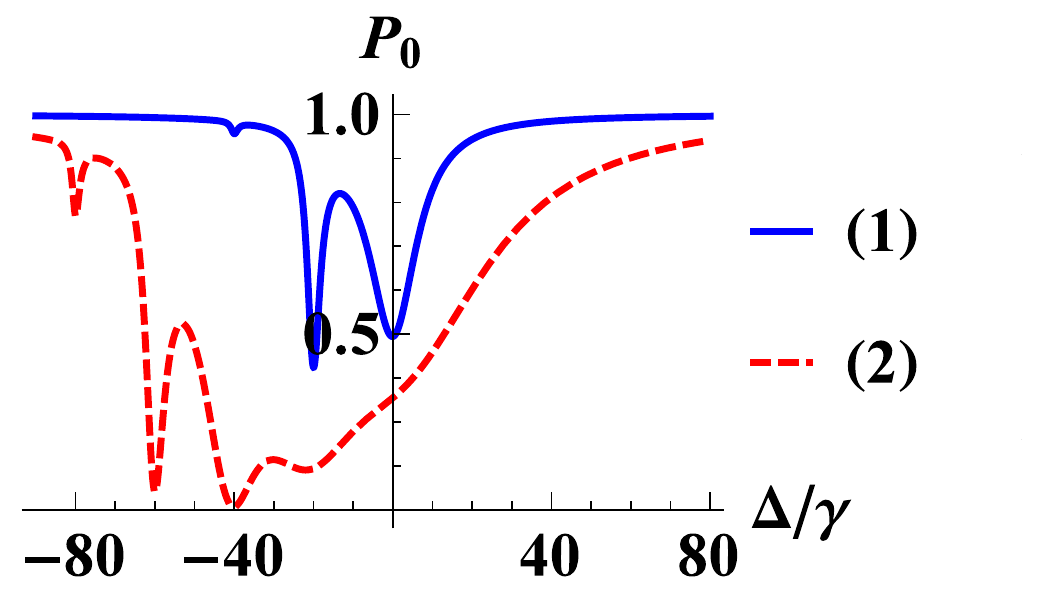}}%
\qquad
\subfloat[][\label{pdelta3b}]{\includegraphics[scale=0.35]{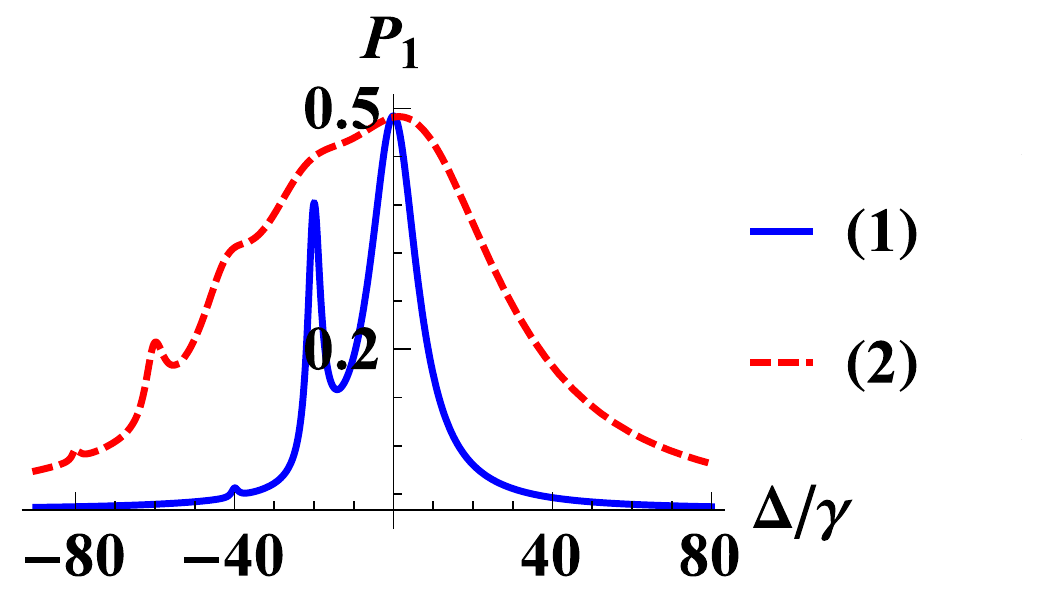}}
\vfill
\subfloat[][\label{pdelta3c}]{\includegraphics[scale=0.35]{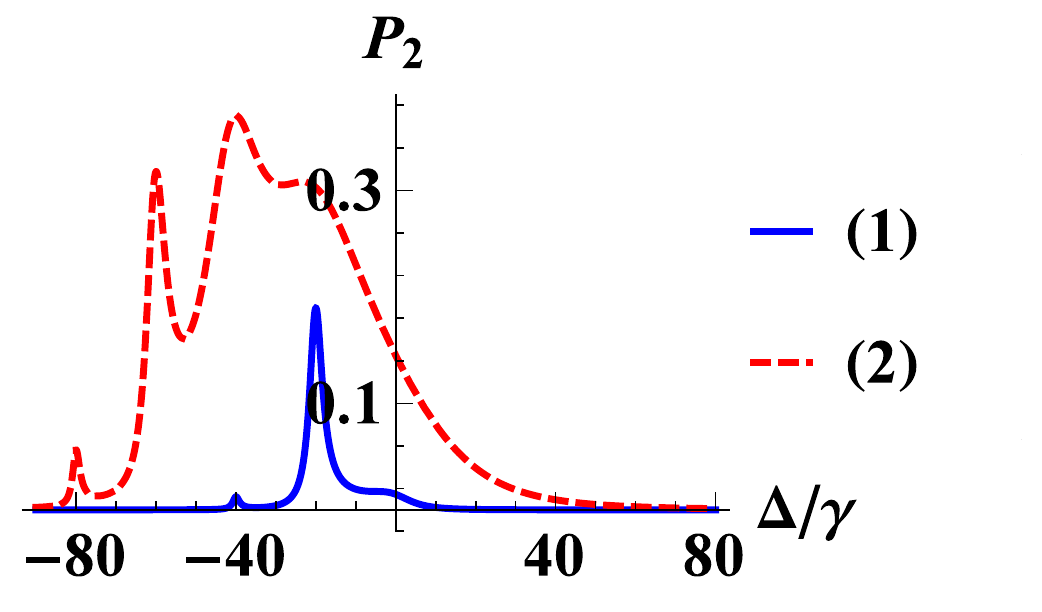}}
\qquad
\subfloat[][\label{pdelta3d}]{\includegraphics[scale=0.35]{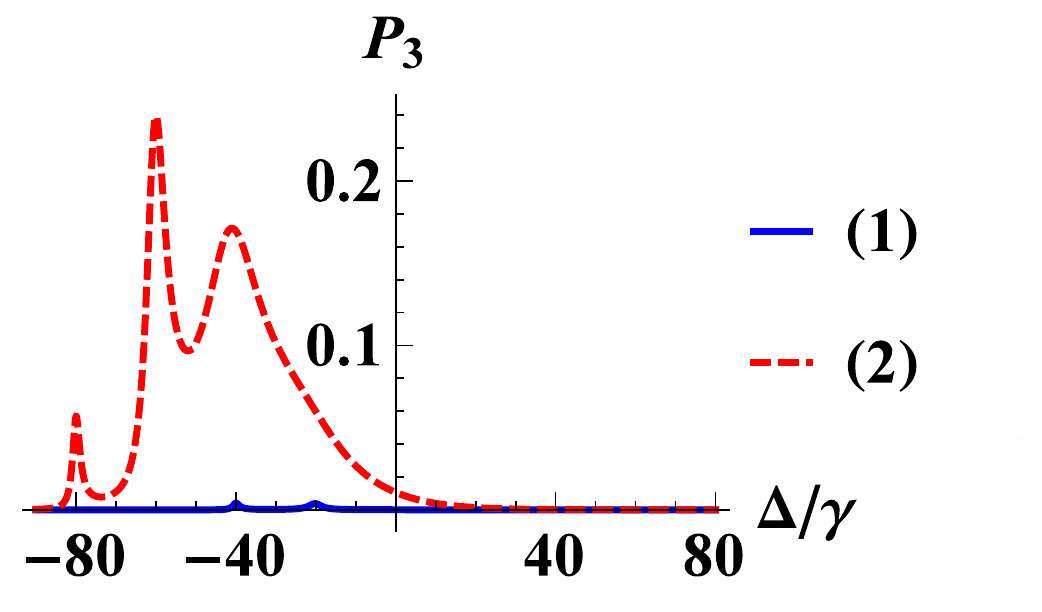}}
\caption{(Color online) Populations of photon numbers   in dependence of the detuning. The parameter are: $\chi/\gamma=20$;  $\Omega/\gamma=5$, curves (1) and $\Omega/\gamma=20$, curves  (2).}
\label{pdelta3}
\end{figure}

In this subsection, we consider excitations of KNR on the base of Eqs. (\ref{phdis}) and (\ref{quant}), assuming the quantum regime, but not the formation of well-resolved  energy levels of KNR.  At first,  two-dimensional plots of quantum mechanical mean photon numbers are depicted on Fig. \ref{pn1}  in dependence of the detuning values $\Delta/\gamma$ and  the driving amplitude $\Omega/\gamma$. Comparing Fig. \ref{pn1a} and Fig. \ref{pn1b} we conclude that  the maximum values of the mean photon numbers in the resonator are essentially small for  large value of the ratio $\chi/\gamma=20$ in comparison with the case of $\chi/\gamma=2$. This means that a high nonlinearity with respect to dissipation leads to formation of a strong quantum operational regime.

 Next, we analyze the photon number distribution in dependence of the detuning, for the case of a moderate nonlinearity, $\chi/\gamma=2$ . In this case the driving field is simultaneously resonant with many transitions between consecutive oscillatory levels. The results of calculations on the base of Eq. (\ref{phdis})  are depicted on Fig. \ref{pn3d2a} as two-dimensional plots of the Fock state populations. These plots illustrate the formation of  Poissonian-type distributions from the thermal distribution, if the frequency of the driving field is tuned around the frequency of KNR. Nevertheless, these distributions are narrow compared to the Poissonian distributions for a higher amplitude of the driving field. Photon-number distribution functions are plotted on Fig. \ref{pn3d2b} for  different values of the detuning.

\subsection{Selective resonant excitations of oscillatory states}\label{SecPD2}

\begin{figure}%
\subfloat[][\label{pkhi4a}]{\includegraphics[scale=0.6]{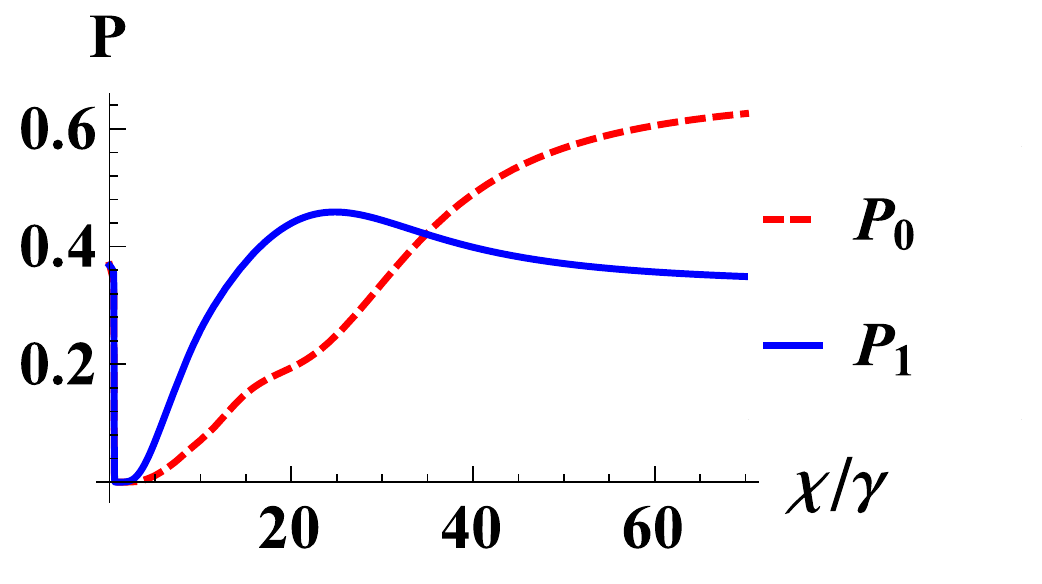}}
\qquad
\subfloat[][\label{pkhi4b}]{\includegraphics[scale=0.6]{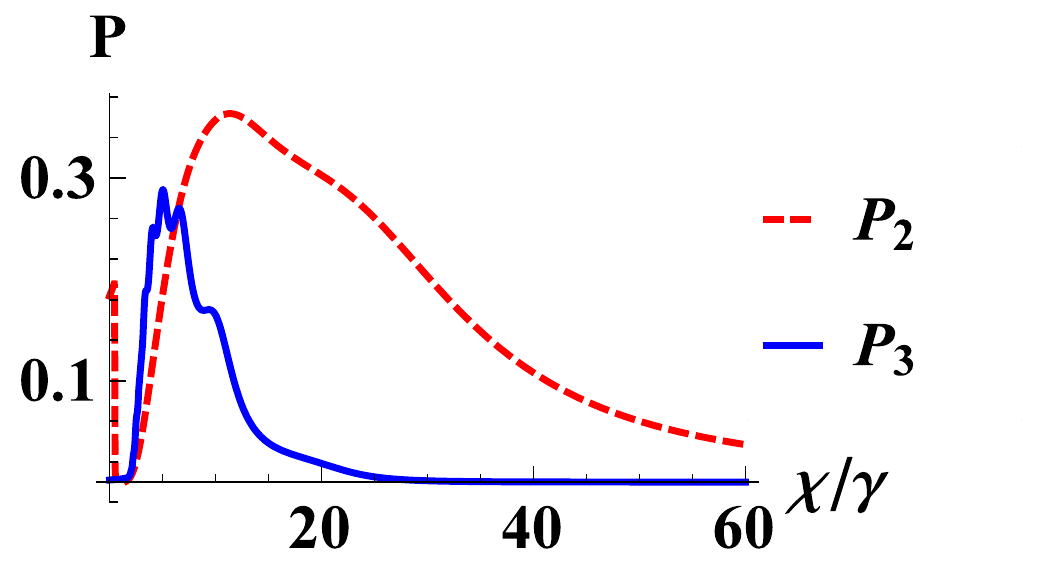}}
\caption{(Color online)Photon probability distributions $P_0$ and $P_1$ (a), $P_2$ and $P_3$ (b) in dependence of the nonlinearity. The parameters used are: $\Delta/\gamma=-20$, $\Omega/\gamma=20$.}
\label{pkhi4}
\end{figure}

\begin{figure}%
\begin{center}
\subfloat[][\label{pomega5a}]{\includegraphics[scale=0.6]{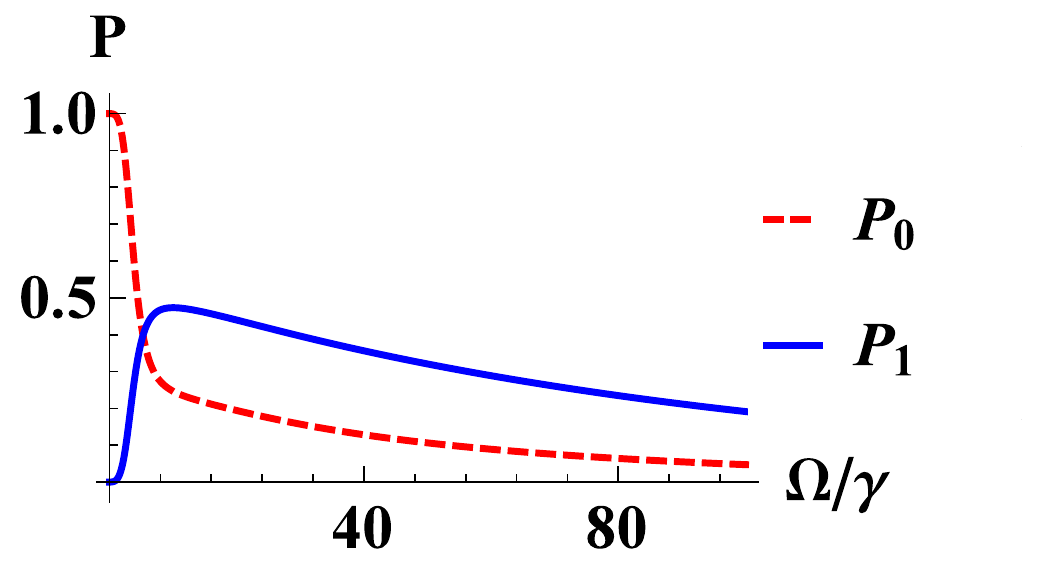}}%
\qquad
\subfloat[][\label{pomega5b}]{\includegraphics[scale=0.6]{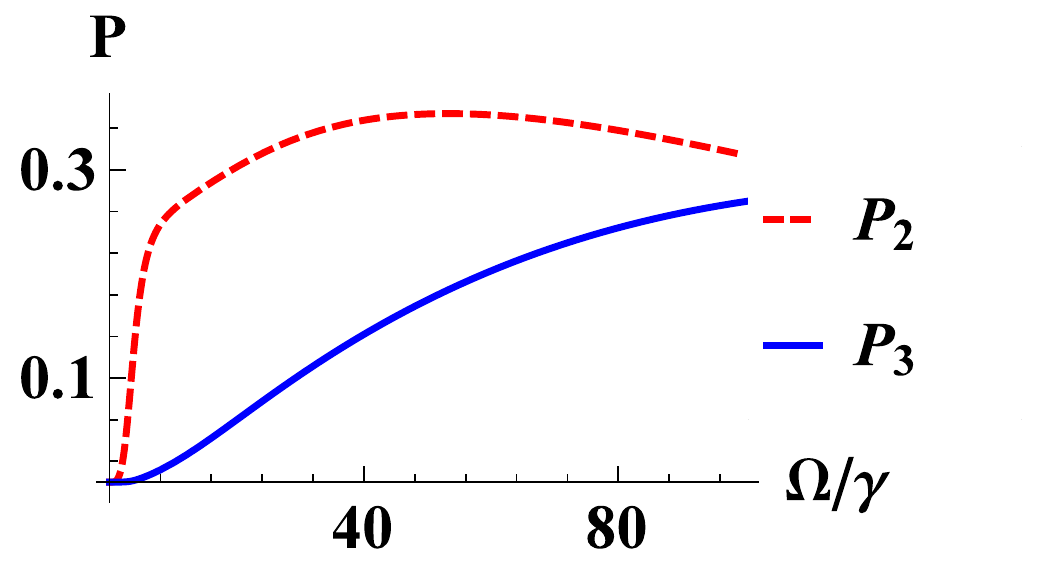}}
\caption{(Color online)Photon probability distributions $P_0$ and $P_1$ (a), $P_2$ and $P_3$ (b) in dependence of the external force amplitude. The parameters are as follows: $\Delta/\gamma=-20$, $\chi/\gamma=20$.}
\label{pomega5}
\end{center}
\end{figure}

In this subsection we discuss the case of very strong nonlinearities for which energy levels of KNR are well resolved, and hence selective excitations of the oscillatory mode take place. Figure \ref{pdelta3} illustrates the dependence of $P_0, P_1, P_2, P_3$ populations of the photon-number states $|n\rangle$ on the detuning $\Delta/\gamma$ for the case of very strong nonlinearity, $\chi/\gamma=20$. As we see, selective  resonance  excitations of the Fock states $|1\rangle$, $|2\rangle$,  $|3\rangle$  take place for both cases of driving field:  $\Omega/\gamma=5$ and  $\Omega/\gamma=20$. Indeed, the populations $ P_1$ depicted on Fig. \ref{pdelta3b} display a maximum of 0.5 at $\Delta=0$, that corresponds to the resonant one-photon transition $|0\rangle \rightarrow |1\rangle$ at the frequency $\omega = \omega_{0}$.
The other maximum $P_1 = 0.38$, for $\Omega/\gamma=5$, corresponds to the population of state $|1\rangle$ through the Raman process with energy conversation $E_0+2\omega_2= \omega_k+E_1$. This process involves a two-photon excitation of the Fock state $|2\rangle$, (transition $|0\rangle \rightarrow  |2\rangle$ at the frequency of the pump field $\omega_2=\omega_0+ \chi$), and the decay $|2\rangle \rightarrow |1\rangle$ at the frequency $\omega_k= \omega_0+2\chi $.

\begin{figure}%
\subfloat[][\label{3dkhia}]{\includegraphics[scale=0.37]{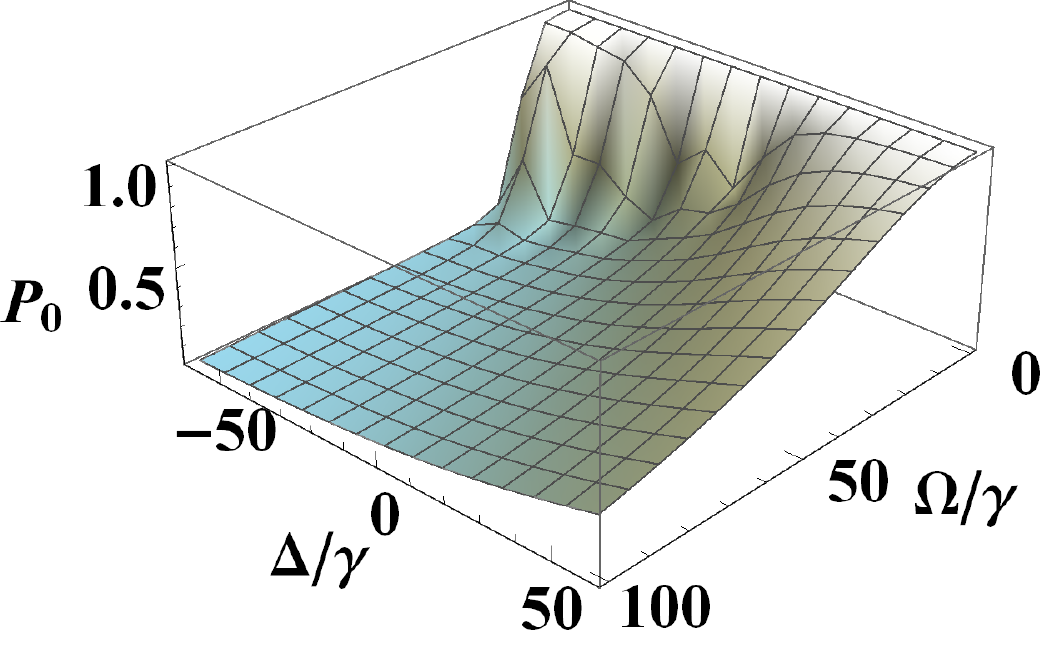}}
\qquad
\subfloat[][\label{3dkhib}]{\includegraphics[scale=0.37]{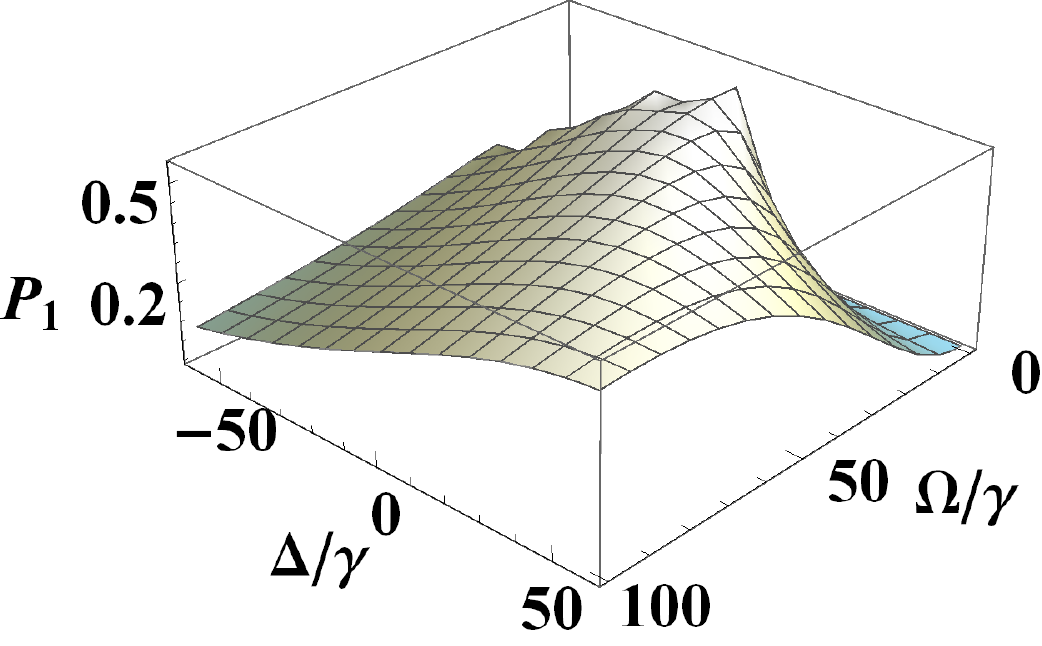}}
\vfill
\subfloat[][\label{3dkhic}]{\includegraphics[scale=0.37]{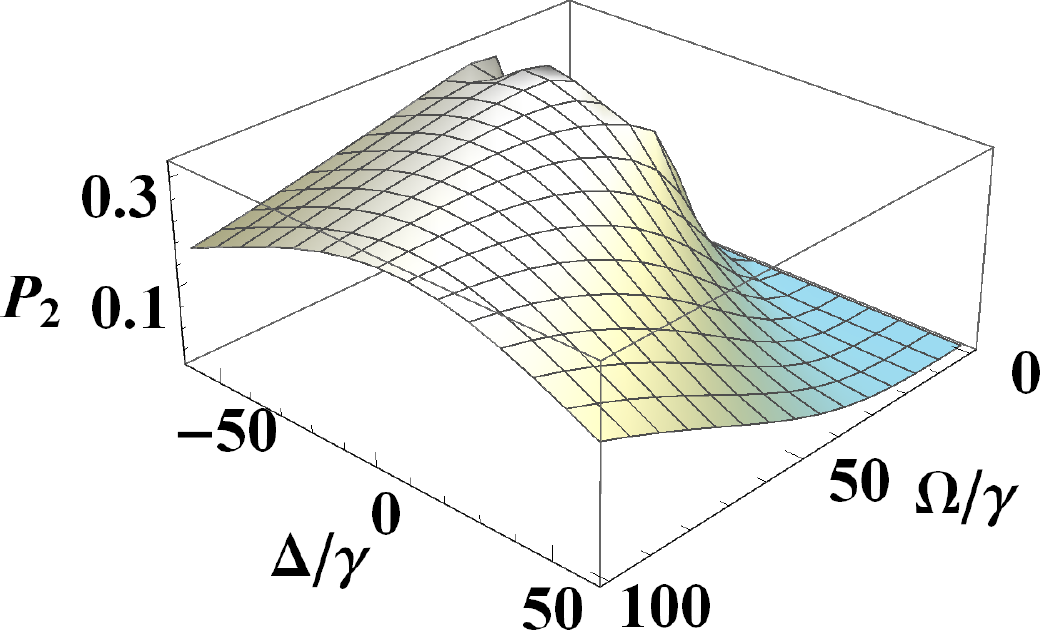}}
\qquad
\subfloat[][\label{3dkhid}]{\includegraphics[scale=0.37]{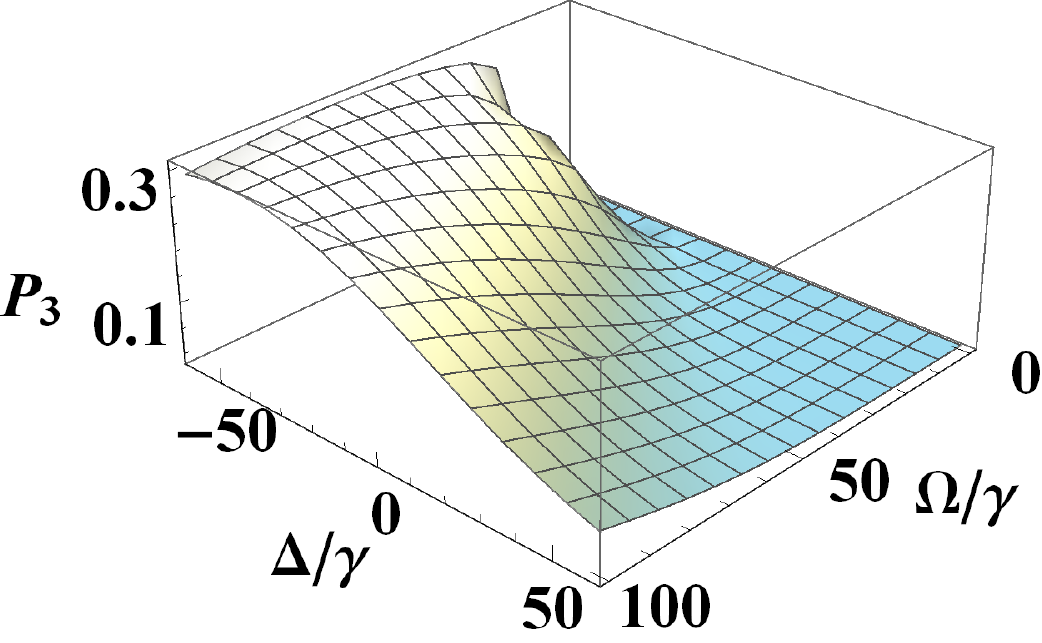}}
\caption{Photon probability distributions $P_0$ (a), $P_1$ (b), $P_2$ (c) and $P_3$ (d) in dependence on the detuning and the field amplitude. The nonlinearity parameter is: $\chi/\gamma=20$.}
\label{3dkhi}
\end{figure}

 As we see on Fig. \ref{pdelta3c}, for a comparatively weak driving field,  $\Omega/\gamma=5$, the population $P_2$  only reaches the value 0.2  due to a two-photon resonant excitation  $|0\rangle \rightarrow |2\rangle$ at the frequency  $\omega_2=\omega_0+ \chi$. In the range of strong excitations the role of multi-photon processes increases. We observe that these populations display peaked structures and the maximum values are realized for definite parameters of the detuning in accordance with resonant frequencies. Indeed, for the case of a strong field,  $\Omega/\gamma=20$,  the Raman processes of state $|2\rangle$ become effective.  Particularly,  the peaks at $\Delta/\gamma=-40$ and $\Delta/\gamma=-60$ correspond to the Raman processes through the states $|3\rangle$  at the frequency $\omega_3=\omega_0 + 2\chi$ and $|4\rangle$ at $\omega_4=\omega_0 + 3\chi$  , respectively. As we see on Fig. \ref{pdelta3d}, for weak excitation,  $\Omega/\gamma=5$ (1), the population  $P_3$ is negligible for all values of the detuning, while  for the case of a strong excitation,  $\Omega/\gamma=20$ (2), a selective population takes place.

In general, from analytically obtained results and the numerical analysis we can conclude that low-energetic states  $|1\rangle$,  $|2\rangle$ and  $|3\rangle$ can be selectively excited with high efficiency in case of  very strong nonlinearities. Nevertheless, we strongly demonstrate  that values of  Fock state populations  are  limited in the cw steady-state regime of KNR.  Indeed, on Fig. \ref{pdelta3b} the population $P_1$  is limited by the value 0.5, and the populations  $ P_2$ and $ P_3$ are less than $ P_1$.  This effect of
selective excitations can be interpreted as blockade of two or more photons in KNR by production of a single-photon Fock state. Thus, observation of these limits on the populations  restricts the  possibility to observe photon blockade effects  in an over transient steady-state regime.

 We demonstrate this fact on Fig. \ref{pkhi4}, showing  the dependence of populations from the parameters of nonlinearity at a fixed detuning value $\Delta/\gamma=-20$ and an external field amplitude $\Omega/\gamma=20$. This regime involves two-photon resonance excitation of KNR for $\chi/\gamma=20$. It can be observed that effective excitations of the oscillatory mode take place only for particular values of the nonlinearity parameter, as a result the values of populations are limited, particularly  $P_1$ is limited by 0.5, while the populations  $P_2$ and  $P_3$ are limited by the values 0.35 and 0.3 correspondingly. Analogously, the populations versus  the amplitude of driving field depicted on Fig. \ref{pomega5} demonstrate limited behavior in steady-state regimes as well. However the behavior of populations of $P_2$ and $P_3$ depicted on Fig. \ref{pomega5b} differ from $P_0$ and $P_1$ (Fig. \ref{pomega5a}). The complete image of  the populations $P_0$, $P_1$, $P_2$ and $P_3$ for the nonlinearity parameter $\chi/\gamma=20$ (case of selective excitations), in dependence of the detuning $\Delta/\gamma$ and the amplitude of the driving field $\Omega/\gamma$ are presented on Fig. \ref{3dkhi}.

\subsection{Quantum statistics and Wigner functions}\label{SecPD3}

\begin{figure}%
\subfloat[][\label{g7a}]{\includegraphics[scale=0.6]{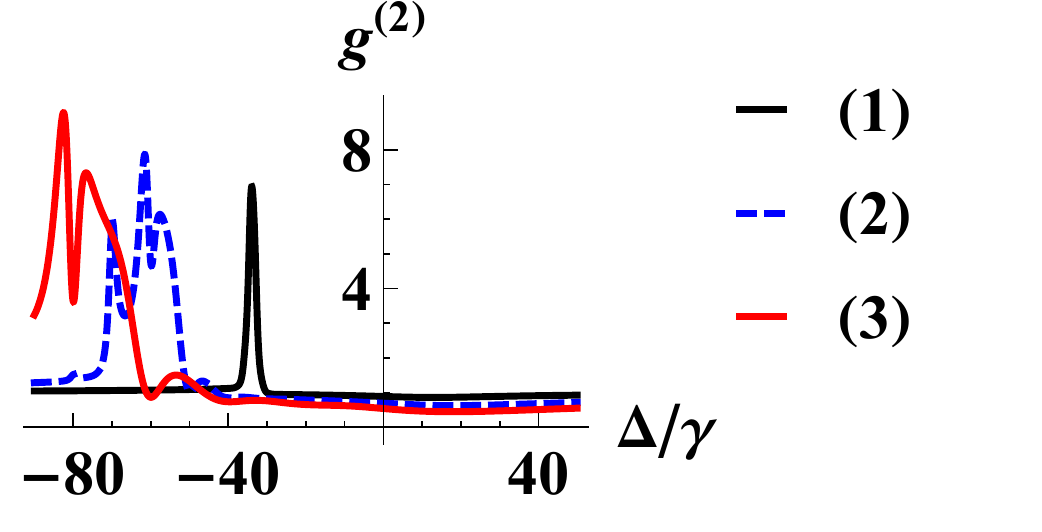}}%
\qquad
\subfloat[][\label{n7b}]{\includegraphics[scale=0.6]{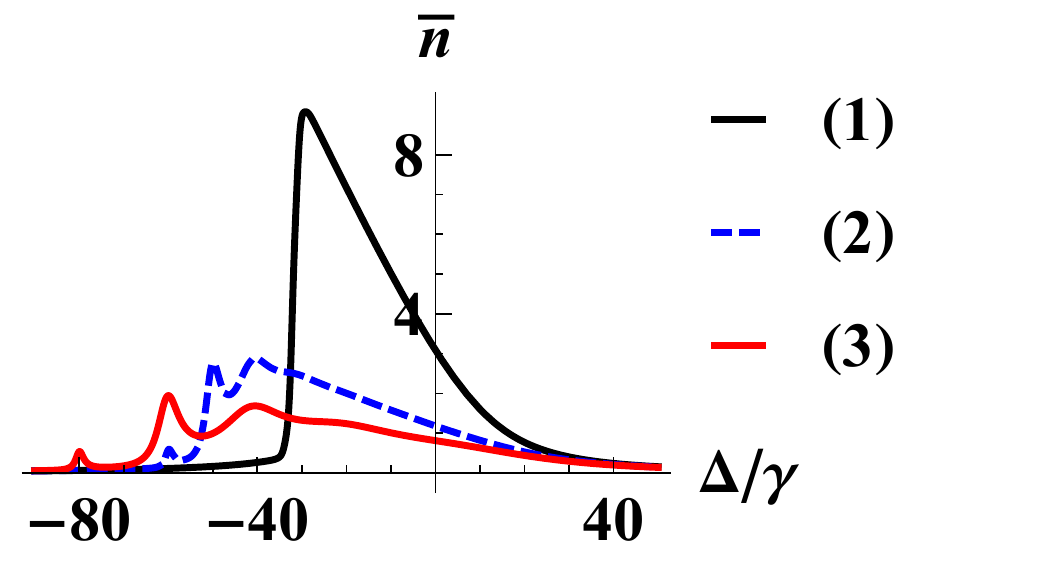}}
\caption{The correlation function (a) and the mean excitation number (b) in dependence on the detuning for the amplitude of the driving field $\Omega/\gamma=20$.  The nonlinearity parameter is: $\chi/\gamma=2$ (1), $\chi/\gamma=10$ (2), $\chi/\gamma=20$ (3).}
\label{gn7}
\end{figure}
In the end of this section, we turn to investigation of photon statistics on the  base of the normalized second-order correlation function (\ref{g_init}) for zero delay time.  The correlation function describes the ratio of the simultaneously emitted photon pair number to the product of the independently emitted photon number, and, on the other hand, is expressed through the variance of the photon number fluctuations as $\langle (\Delta n)^2\rangle = \langle n\rangle + \langle n\rangle ^2(g^{(2)}-1)$. Thus, the condition $g^{(2)}<1$ corresponds to the sub-Poissonian statistics $\langle (\Delta n)^2\rangle < \langle n\rangle$ , the condition $g^{2}(0) = 1$ corresponds to the Poissonian statistics and  the condition $g^{2}(0) > 1$ to  the super-Poissonian statistics. The results for  $g^{2}(0)$ and the mean photon numbers are depicted on Fig. \ref{gn7} for three values of the nonlinearity  $\chi/\gamma=2$,  $\chi/\gamma=10$,  $\chi/\gamma=20$. As we see, for the regime of non-resonant excitations, ($\chi/\gamma=2$), the correlation function approximately equals to unity for a large range of the detuning. This result is in accordance with the results depicted on Fig. \ref{pn3d2} and reflects the Poissonian statistics of KNR. Note, the correlation function displays a peak around the value of detuning $\Delta/\gamma=-35$  in the critical or threshold region of photons production. An analogous result on critical growth of the correlation functions in the region of the generation threshold has been demonstrated for the non-degenerate parametric oscillator \cite{crit}. For the regimes of very strong nonlinearities the correlation function is less than unity ($g^{(2)}<1$), i.e. the mode displays sub-Poissonian non-classical statistics for a large range of detuning values from $\Delta/\gamma=-40$ to $\Delta/\gamma=40$ due to selective excitations of KNR.

\begin{figure}%
\subfloat[][\label{3dw8a}]{\includegraphics[scale=0.6]{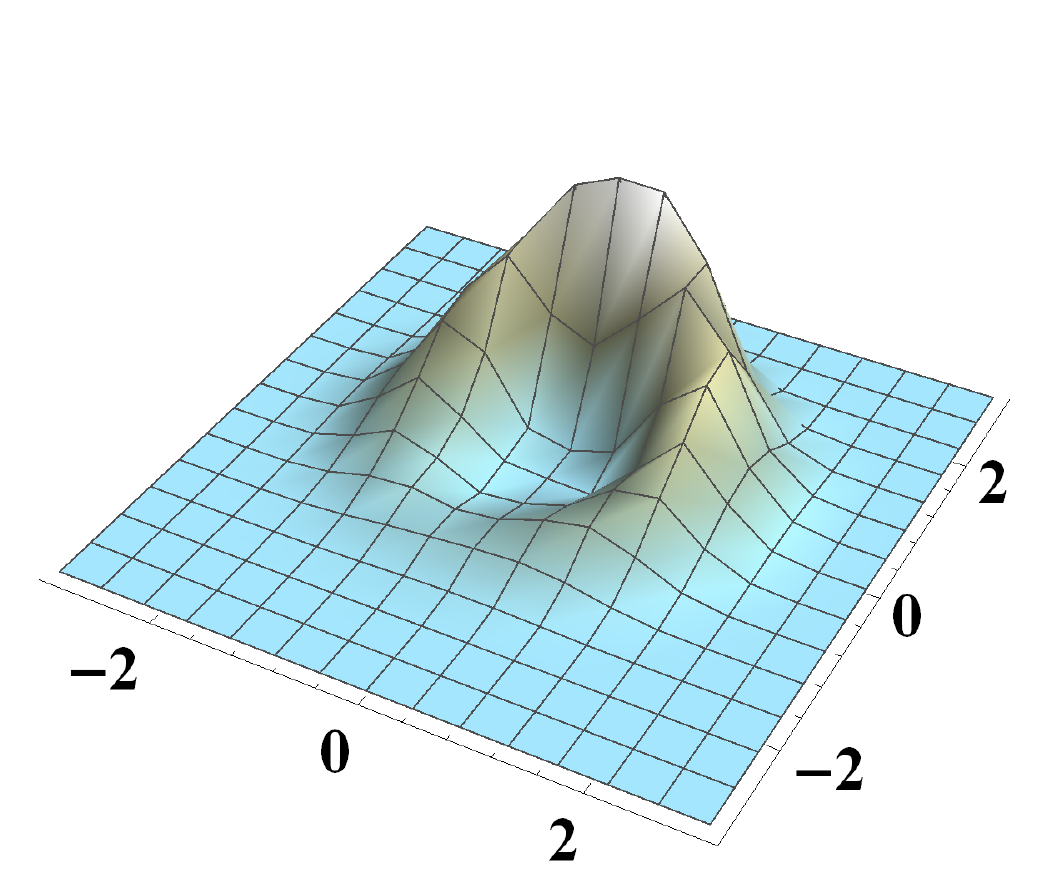}}%
\qquad
\subfloat[][\label{pn8b}]{\includegraphics[scale=0.6]{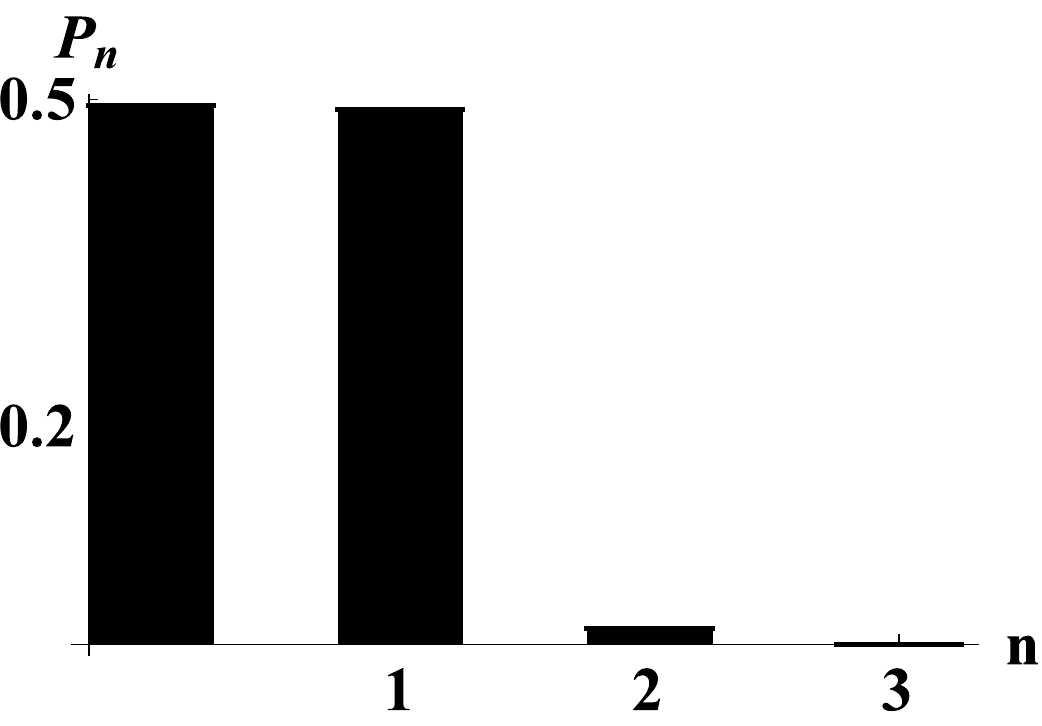}}
\caption{The Wigner function (a) and  the distribution of photon number states (b) for the parameters $\Omega/\gamma=5$, $\Delta/\gamma=0$, $\chi/\gamma=20$.}
\label{wigner8}
\end{figure}

\begin{figure}%
\subfloat[][\label{3dw9a}]{\includegraphics[scale=0.6]{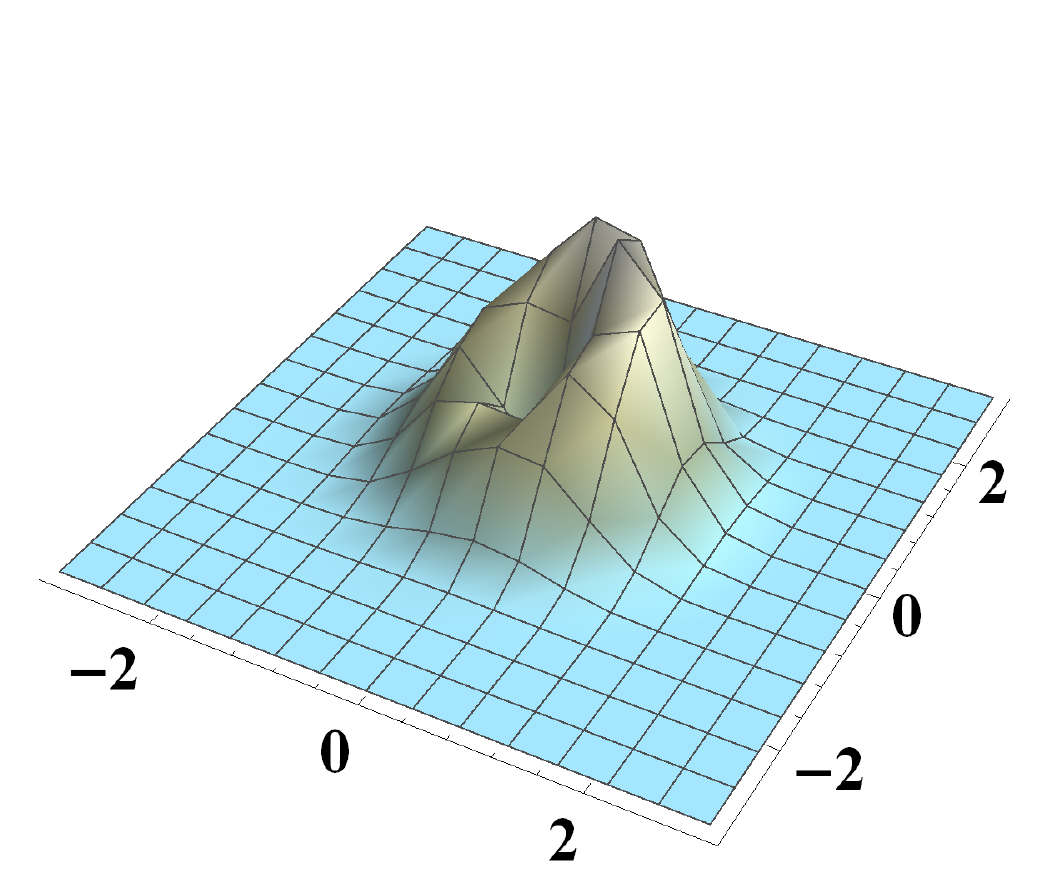}}%
\qquad
\subfloat[][\label{pn9b}]{\includegraphics[scale=0.6]{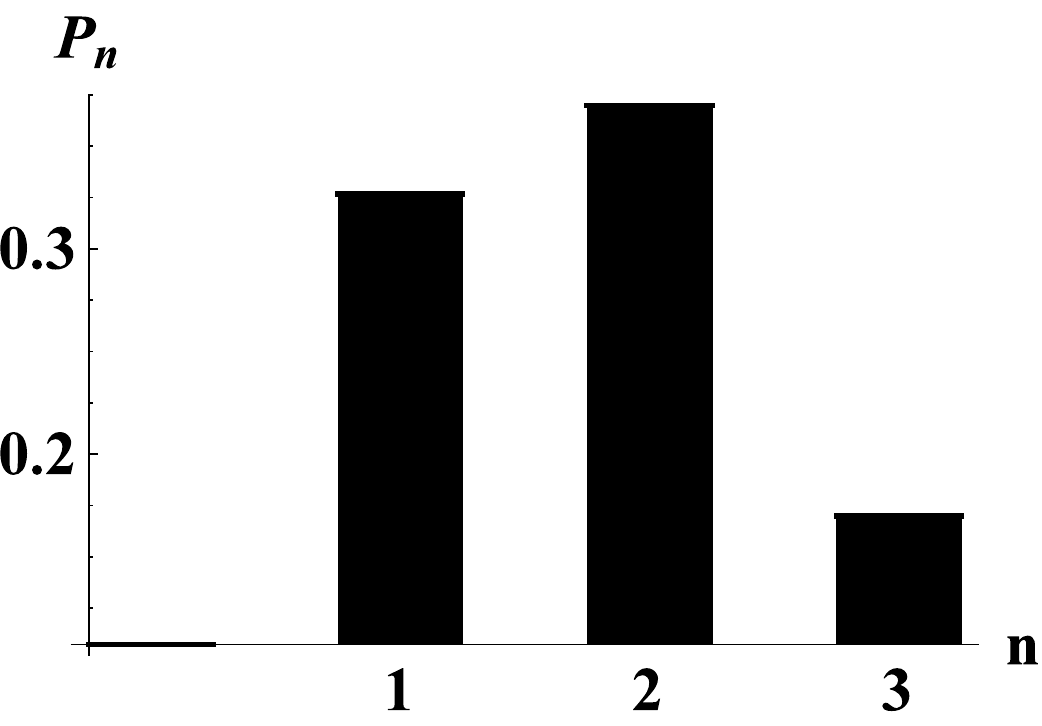}}
\caption{The Wigner function (a) and  the distribution of photon number states (b) for the  parameters $\Omega/\gamma=20$, $\Delta/\gamma=-40$, $\chi/\gamma=20$.}
\label{wigner9}
\end{figure}

At the end of this subsection we turn to discussion of phase-space properties of KNR in regimes of selective excitations. Typical results for the Wigner function of the cavity mode and the corresponding distributions of photon number states  are depicted on  Fig. \ref{wigner8} and Fig. \ref{wigner9}. At first, we consider the case of the parameters $\Omega/\gamma=5$, $\Delta/\gamma=0$, and $\chi/\gamma=20$. This regime interests us in terms that only vacuum and single photon states are allowed here and the probabilities of both states equal to (~0.5) (see, Fig. \ref{pdelta3}). The results of calculations (4) and (11) are illustrated on Figs. \ref{3dw8a} and \ref{pn8b}.  As the Wigner function is positive overall the phase-space, so  KNR driven by a monochromatic force can not display any quantum-interference pattern.  Nevertheless, the Wigner function displays non-classical photon-number effects in strong quantum regimes. On Fig. \ref{wigner9} the Wigner function and the photon number probabilities are illustrated for the case of a strong field  $\Omega/\gamma=20$ and for the parameters $\Delta/\gamma=-40$ and $\chi/\gamma=20$. The probability of the vacuum state is neglectably low in this case (see, Fig. \ref{pdelta3a}). In this regime only one-photon and two-photon Fock states are effectively  observed and the Wigner function displays a specific form in the phase-space. 
Thus, Fig.9 depicts distributions of n-photon states showing considerable contributions from the one- two- and three-photon states. The similar situation was already discussed, although for the perturbation regime \cite{leon}.

\section{Conclusion}\label{Conclusion}

In conclusion, we  have investigated  various regimes of mode  excitations in KNR driven by a continuous wave field for  strong  Kerr nonlinearities.  The results  have been obtained on the base of exact quantum solution of the Fokker-Planck equation in the complex $P$ representation. This approach allows non-perturbative consideration of KNR for various quantum operational ranges  including   cascaded multiphoton  processes between oscillatory states stimulated by a strong driving field and for quantum noise of arbitrary strength.  The selective excitations of the resonator mode are realized  in the cases of well resolved energy spectra of an anharmonic oscillator by tuning the frequency of the driving field for wide ranges of the system parameters, the nonlinearity constant, the detuning and the amplitude of the driving field. On the base of this approach the mean photon numbers have been obtained as well as the photon statistics of the resonator mode has been analyzed using the probability distributions of photons, the second-order photon correlation functions and the Wigner functions. It has been shown that for moderate nonlinearities photon state distributions are close to Poissonian distributions, while for strong nonlinearities selective excitations of photon number states may arise. Particularly, we have demonstrated  the limited behavior of photon number-state populations in steady-state regimes due to concurrency of cascaded processes and decoherence effects. We observed that the population of one-photon state is limited by the value 0.5 while the population of two-photon state is limited by 0.35 for all ranges of the parameters. This conclusion  restricts the possibilities for realization  of the photon blockade due to cw excitations in over transient steady-state regimes.

We thank Prof. I.A. Shelykh  and Prof. O.V. Kibis for the discussions we had on the subject. We acknowledge support of the Armenian State Committee of Science, the Project No.15T-1C052 and EC for the RISE, Project CoExAN GA644076. 

All authors contributed equally in the work presented in this paper.

\end{document}